%% file: kant-spg-solver.tex
\documentclass[submission,copyright,creativecommons]{eptcs}
\providecommand{\event}{GRAPHITE 2014} 
\usepackage{breakurl}             
\usepackage{prelude}

\begin{document}

\title{Generating and Solving Symbolic Parity Games}
\author{
\begin{tabular}{c@{\qquad\qquad}c}
Gijs Kant\thanks{Gijs Kant is sponsored by the NWO under grant number 612.000.937 (VOCHS).}
& Jaco van de Pol\\
\texttt{\footnotesize kant@cs.utwente.nl}  & \texttt{\footnotesize vdpol@cs.utwente.nl}
\end{tabular}
\medskip
\\{\footnotesize Formal Methods \& Tools}
\\{\footnotesize University of Twente, Enschede, The Netherlands}
}
\def\titlerunning{Generating and Solving Symbolic Parity Games}
\def\authorrunning{G. Kant \& J.C. van de Pol}

\maketitle

\begin{abstract}
We present a new tool for verification of modal \MUCALC formulae for process specifications,
based on symbolic parity games.
It enhances an existing method, that first encodes the problem to a Parameterised Boolean Equation System (PBES) and then instantiates the PBES to a parity game.
We improved the translation from specification to PBES to preserve the structure of the specification in the PBES,
we extended \LTSMIN to instantiate PBESs to symbolic parity games, and implemented the recursive parity game solving algorithm by Zielonka for symbolic parity games. 
We use Multi-valued Decision Diagrams (MDDs) to represent sets and
relations, thus enabling the tools to deal with very large systems.
The transition relation is partitioned based on the structure of the specification, which allows for efficient manipulation of the MDDs.
We performed two case studies on modular specifications, that demonstrate that the new method has better time and memory performance than existing PBES based tools and can be faster (but slightly less memory efficient) than the symbolic model checker \NuSMV.
\end{abstract}


\input{01-intro.tex}
\input{02-background.tex}
\input{03-lps2pbes.tex}
\input{04-spg-generation.tex}
\input{05-spg-solver.tex}

\input{06-experiments.tex}
\input{07-conclusions.tex}

\bibliographystyle{eptcs}
\bibliography{bibliography/kant-phdproject}

\end{document}

%% file: 01-intro.tex
\section{Introduction}

When verifying large systems or modelling large games with, say, billions or even trillions of states, 
datastructures are needed that can represent such large numbers of states, e.g., 
Multi-valued Decision Diagrams (MDDs). 
We have developed a tool that enables verification of modal \MUCALC formulae for process algebraic specifications
using MDDs.
The tool, called \texttt{pbes2lts-sym}, is now part of \LTSMIN, a toolset for high performance verification that is
language-independent \cite{blom2010:ltsmin}\footnote{Available from \url{http://fmt.cs.utwente.nl/tools/ltsmin} (Open Source).}. The tool can deal with very large state spaces,
provided that the transition relation of the modelled system can be partitioned into relatively independent groups.
\LTSMIN is used in several application domains, including verification of railway safety systems. In this paper a case study is included where the presented method is applied to analysis of the control software used in the Large Hadron Collider at CERN.

\begin{figure}[tpb]
\centering
\scalebox{0.8}{
\input{images/lps2pg.tikz}
}
\caption{Overview.}
\label{fig:overview}
\vspace{-10pt}
\end{figure}
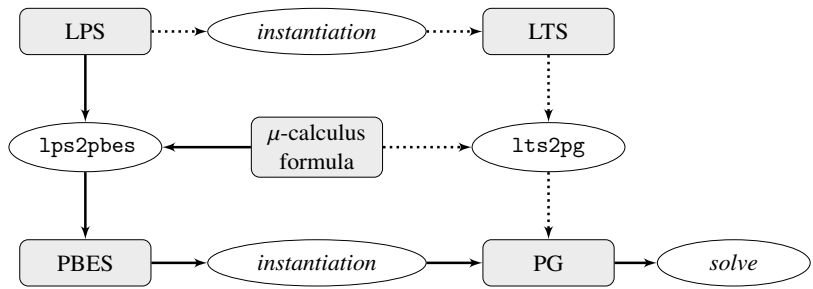

An established method for verification of \MUCALC is translation of the problem to a Parity Game (PG) and then solving the game.
Our starting point is a Linear Process Specification (LPS), specified in the process algebraic language \MCRLTWO.
A possible translation of the verification problem to a parity game is shown as the dotted route through Fig.~\ref{fig:overview}: 
first instantiating the LPS to a Labelled Transition System (LTS) and then translating satisfaction of a formula by the LTS to a parity game.
We follow the more symbolic route, taken in the \MCRLTWO toolset.\footnote{See \url{http://mcrl2.org}.} 
This verification approach corresponds to the solid line route in Fig.~\ref{fig:overview}.
The problem is first translated to a Parameterised Boolean Equation System (PBES) \cite{groote2005:parameterised}, a sequence of Boolean fixpoint equations with data variables, of which the solution is 
\ttrue if and only if the specification satisfies the formula.
The PBES is instantiated to a parity game with the same solution.
An advantage of the second approach is that the intermediate step of generating the LTS, which can be rather large, is not needed. Furthermore, property-specific reduction techniques can be applied to the PBES, which would result in a smaller parity game.

Parity games are two player games, represented by a game graph where the nodes
represent the states of the game and the edges the possible moves, and each node belongs to one of the players (representing `and' and `or') and has a priority. 
Solving a parity game (locally) means determining if a winning strategy from the initial state exists for one of the players.
The concepts that are used will be briefly explained in Section~\ref{section:background}.

When the system is very large, also the parity game that encodes satisfaction of a formula for that system can become very large.
Therefore we need efficient data structures and algorithms to generate and solve the game.

In earlier work \cite{kant2012:efficient}, we presented an early version of the \texttt{pbes2lts-sym} tool for generating parity games from PBESs. 
We use symbolic parity games, in which MDDs are used to represent sets of states 
and 
the relations encoding the moves, partitioned in transition groups.
Other existing tools for solving PBESs, available in the \MCRLTWO toolset, use an explicit state representation, which severely
limits the size of system that can be verified.
The \texttt{pbes2lts-sym} tool is based on technology for generating symbolic state spaces, as is the tool \texttt{lps2lts-sym} for 
generating the state space for an LPS, which is described extensively in \cite{blom2008:symbolic}.
In \texttt{lps2lts-sym}, states are represented as vectors of values and are stored in an MDD.
In order to be efficient, it is required that the specification is modular, i.e., that
the transition relation can be partitioned into transition groups and that each of these groups depend on and influence only a small part of the state vector.
This locality of transitions is expressed in a dependency matrix.
When this matrix is sparse, the relations can be represented very compactly and applied to sets of states efficiently.
Because of that, the structure of a system is more important than its size (we consider locality of transitions to be a structural property op the system).
For \texttt{pbes2lts-sym}, the structure of the PBES is equally important.
The generation of symbolic parity games is described in Section~\ref{section:spg-generation}.

In the existing translation from specification to PBES, as described in \cite{groote2005:modelchecking} and available in the \MCRLTWO toolset, the structure of the process specification is not 
explicitly visible in the generated PBES.
One equation is generated per propositional variable in the formula, disregarding the structure of the system.
This makes it impossible to choose a good partition of the transition relation for generating a parity game.
In the previous version of \texttt{pbes2lts-sym}, we tried to guess a partition by splitting conjunctive equations in conjuncts and disjunctive equations in disjuncts.
That way symbolic data structures could be used, but far from optimal 
since it disregards the original process structure, hindering efficient MDD manipulation.
The importance of the partitioned transition relation in symbolic verification is well 
known in the literature and the basis of the success of tools like \NuSMV \cite{cimatti2002:nusmv2} and techniques like saturation \cite{ciardo2001:saturation}.

In this article we propose a modified translation from LPS to PBES
to preserve the structure such that the \texttt{pbes2lts-sym} tool can 
base the partitioning on the structure of the specification and can 
benefit in the same way from a sparse dependency matrix as the symbolic tools for generating LTSs.
The new translation to PBESs is presented in Section~\ref{section:lps2pbes}.

The previous version of \texttt{pbes2lts-sym} did only generate parity games, not solve them.
We implemented the recursive algorithm of Zielonka \cite{zielonka1998:infinite} (similar to \cite{bakera2009:solving}) for symbolic parity games. The solver is decribed in Section~\ref{section:spg-solver}.
The combination of generation and solving is now available as part of the \LTSMIN toolset.

We performed two large case studies, presented in Section~\ref{section:experiments},
solving the Connect Four game and verification of control software in use at the
CMS particle detector at CERN.\footnote{Instructions on how to install and use the tools and the
files used in the case studies are available at
\url{http://www.cs.utwente.nl/\~kant/graphite2014/}.}
Both are challenging problems for model checking tools. The new tools show significant
improvement of time and memory performance over existing tools in the \MCRLTWO toolset and \texttt{pbes2lts-sym} with the previous translation from LPS to PBES.
For the Connect Four game we also compared to \NuSMV, where our tool is faster,
but \NuSMV has better memory performance.

%% file: images/lps2pg.tikz
\tikzstyle{textblock} = [rectangle, text width=5em, text centered, minimum height=2em]
\tikzstyle{block} = [rectangle, draw, fill=gray!15, text width=5em, text centered, rounded corners, minimum height=2em]
\tikzstyle{process} = [ellipse, draw, text width=4em, text centered, minimum height=2em]
\tikzstyle{arrow} = [draw, -stealth']
\tikzstyle{transformation} = [draw, -latex', very thick]

\begin{tikzpicture}[node distance=8em, auto, bend angle=50]
  \node [block] (mcrl) {LPS};
  \node [process, below of=mcrl, node distance=5em] (lps2pbes) {\texttt{lps2pbes}};
  \node [block, below of=lps2pbes, node distance=5em] (pbes) {PBES};

  \node [process, right of=mcrl, node distance=10em, text width=6em] (mcrlinst) {\textit{instantiation}};
  \node [process, right of=pbes, node distance=10em, text width=6em] (pbesinst) {\textit{instantiation}};  
  \node [block, below of=mcrlinst, node distance=5em] (mcalc) {$\mu$-calculus formula};
  
  \node [block, right of=mcrlinst, node distance=10em] (lts) {LTS};
  \node [process, below of=lts, node distance=5em] (lts2pg) {\texttt{lts2pg}};  
  \node [block, below of=lts2pg, node distance=5em] (bes) {PG};
  \node [process, right of=bes, node distance=8em, text width=4em] (solvebes) {\itshape solve};

  \path [transformation] (mcrl) -- (lps2pbes);
  \path [transformation] (mcalc) -- (lps2pbes);
  \path [transformation] (lps2pbes) -- (pbes);
  \path [transformation, dotted] (lts) -- (lts2pg);
  \path [transformation, dotted] (mcalc) -- (lts2pg);
  \path [transformation, dotted] (lts2pg) -- (bes);
  \path [transformation, dotted] (mcrl) edge node {} (mcrlinst);
  \path [transformation, dotted] (mcrlinst) edge node {} (lts);
  \path [transformation] (pbes) edge node {} (pbesinst);
  \path [transformation] (pbesinst) edge node {} (bes);
  \path [transformation] (bes) -- (solvebes);
\end{tikzpicture}

%% file: 02-background.tex
\section{Background}\label{section:background}

In this section we will briefly explain Linear Process Specifications,
Modal \MUCALC,
Parameterised Boolean Equation Systems and Parity Games.
More extensive descriptions can be found in, e.g., \cite{groote2001:algebraic} (linear processses), 
\cite{groote2005:parameterised} (PBESs), 
\cite{bradfield2001:modal} and \cite{graedel2002:automata} (parity games).

\subsection{Process algebra}

The \MCRLTWO language is a process algebraic modelling language with 
algebraic data types.
Several analysis techniques are available in the \MCRLTWO toolset, such as simulation, visualisation and model checking.
For analysis purposes a specification is \emph{linearised} to Linear Process Specification (LPS) format. 
Linearisation removes concatenation of actions, parallel composition, hiding, etc.
An LPS consists of a single process with \emph{summands} of the form
``$\sum_{d \oftype D} \ \text{guard} \guards \text{action} \suchthat \text{next state}$''
and an initial state of the process.
The LPS gives rise to a Labelled Transition System (LTS).
The summands model the nondeterministic choice in the system. 

\begin{definition}[Linear Process]\label{def:linear-process}
A \emph{linear process} has the following structure:
\begin{align*}
\proc P(x_p \oftype D_p) & = \sum_{i \in I} \sum_{y \oftype E_i} c_i(x_p, y) 
                            \guards a_i(f_i(x_p,y)) \suchthat P(g_i(x_p,y))
\end{align*}
The specification consists of $m$ summands $i \in I$ with $I = \set{1, \ldots, m}$. Each summand may sum over a data sort $E_i$,
has a guard $c_i$, a parameterised action $a_i$, that is enabled if the guard is satisfied, and the specification of the behaviour after
having executed the action, specified by the recursive definition $P(g_i(x_p,y))$.
In examples the $+$-operator will be used for combining summands instead of the sum notation.
\end{definition}

\begin{example}\label{example:buffer2}
As an example we give a model of a buffer with two cells:
\begin{quote}
\begin{tabular}{l@{}l}
$\proc\ $ & $\procname{Buffer}(q \oftype \container{List}(D)) = $ \\
          & \qquad $\displaystyle\sum_{d \oftype D} \ (\#q < 2) \guards \action{read}(d) \suchthat \procname{Buffer}(q \append d)$ \\
        & \qquad $\ \ \, + \ (q \neq \emptyseq) \guards \action{send}(\head(q)) \suchthat \procname{Buffer}(\tail(q));$ \\
$\init\ $ & $\procname{Buffer}(\emptyseq);$
\end{tabular}
\end{quote}
The process $\procname{Buffer}$ has a data parameter $q$, which is a list of elements from $D$, modelling the contents of the buffer. The size of $q$ is denoted $\#q$.
The process can either read a value $d$ and proceed with $\procname{Buffer}(q \append d)$ (the same process, but with $d$ appended to $q$), 
or send the first element of $q$ and proceed with $\procname{Buffer}(\tail(q))$.
The system is initialised to the $\procname{Buffer}$ process with $q = \emptyseq$, the empty list, as parameter.
\end{example}

\subsection{First order modal \MUCALC}

The first order modal \MUCALC is Hennesy-Milner logic extended with fixpoint operators,
quantifiers and data parameters. Formulae are defined by the grammar:
\[ \phi \Coloneqq   b 
              \mid  \neg b
              \mid  \phi \land \phi
              \mid  \phi \lor \phi
              \mid  \always{\alpha}\phi
              \mid  \possibly{\alpha}\phi  
              \mid  \mathsf{Q} d \oftype D \suchthat \phi
              \mid  \propvar{Z}(e)
              \mid  \sigma \propvar{Z}(x \oftype D \becomes d) \suchthat \phi
\]
where $b$ is a boolean data expression, 
$\mathsf{Q} \in \set{\forall,\exists}$,
$\sigma \in \set{\mu, \nu}$ is a minimal ($\mu$) or maximal ($\nu$) fixpoint operator,
with the restriction that each propositional variable $\propvar{Z}$ occurs positively in $\varphi$ in an equation $\sigma \propvar{Z}(x \oftype D \becomes d) \suchthat \varphi$, i.e.,
within the scope of an even number of negations.
$\alpha$ is some language for specifying predicates on actions.
The class of first order modal \MUCALC formulae is denoted $\mathcal{M}$.
The semantics of formulae interpreted over LTSs is presented in, e.g., \cite{groote2005:modelchecking}.

\subsection{Parameterised Boolean Equation Systems}

Satisfaction of a formula by a linear process specification is first translated to a Parameterised Boolean Equation System (PBES), a system of first order boolean equations. We propose a new translation in
Section~\ref{section:lps2pbes}; here we first define what a PBES is.

\begin{definition}
\emph{Predicate formulae} $\xi$ are defined by the following grammar:
\[ \xi \Coloneqq b \mid \propvar{X}(\vec{e}) \mid \neg \xi \mid \xi \oplus \xi \mid \mathsf{Q} d \oftype D \suchthat \xi \]
where $\oplus \in \set{\land, \lor, \impl}$, $\mathsf{Q} \in \set{\forall, \exists}$, 
$b$ is a data term of sort $\Bool$, 
$\propvar{X} \in \X$ is a predicate variable, 
$d$ is a data variable of sort $D$, and 
$\vec{e}$ is a vector of data terms.
We will call any predicate formula without predicate variables a \emph{simple formula}.
We denote the class of predicate formulae $\PF$.
\end{definition}

\begin{definition}
A \emph{First Order Boolean Equation} is an equation of the form:
$\;
  \sigma \propvar{X}(\vec{d} \oftype D) = \xi 
\;$
where $\sigma \in \set{\mu, \nu}$ is a fixpoint operator,
$\vec{d}$ is a vector of data variables of sort $D$, and
$\xi$ is a predicate formula.
The class of first order boolean equations is denoted $\mathcal{E}$.
\end{definition}

\begin{definition}
A \emph{Parameterised Boolean Equation System (PBES)} is a sequence of First Order Boolean Equations:
$\; 
   \eqsys = (\sigma_1 \propvar{X}_1(\vec{d_1} \oftype D_1) = \xi_1) 
   \eqsep \dotsc 
   \eqsep (\sigma_n \propvar{X}_n(\vec{d_n} \oftype D_n) = \xi_n)
$
\end{definition}

The semantics and solution of PBESs are described in, e.g., \cite{groote2005:parameterised}.
The order of the equations does matter.


\subsection{Parity games}\label{section:paritygames}

A \emph{parity game} 
is a game between two players, player \Eloise (also called $\exists$loise or player \emph{even}) and player \Abelard (also called $\forall$belard or player \emph{odd}), where
each player owns a set of places. On one place a token is placed that can be moved 
by the owner of the place to an adjacent place.
The parity game is represented as a graph. We borrow notation from \cite{bradfield2001:modal} and \cite{mazala2002:infinite}.

\begin{definition}[Parity Game]
A \emph{parity game} is a graph $\G = \tuple{V, E, \VEloise, \VAbelard, v_I, \Omega}$, with
\begin{itemize}\noitemsep
\item $V$ the set of vertices (nodes or places or states);
\item $E \oftype V \times V$ the set of transitions;
\item $V_p \subseteq V$ the set of places owned by player $p$, for $p \in \set{\Eloise, \Abelard}$,
	with $\VEloise \cup \VAbelard = V$ and $\VEloise \cap \VAbelard = \emptyset$;
\item $v_I \in V$ the initial state of the game; 
\item $\Omega \oftype V \to \Nat$ assigns a priority $\Omega(v)$ to each vertex $v \in V$.
\end{itemize}
\end{definition}

The vertices in the graph represent the instantiated variables from the equation system. 
The edges represent possible moves of the token (initially placed on $v_I$) and encode dependencies between variables.
In the parity game, player \Eloise owns the vertices that represent disjunctions, player \Abelard the vertices that represent conjunctions.

Player \Eloise is the winner of a play $\pi$ if %
$\pi$ is a finite play $v_0 v_1 \cdots v_r \in V^+$ and
$v_r \in \VAbelard$ and no move is possible from $v_r$; or
$\pi$ is an infinite play and $\min(\Inf(\Omega(\pi)))$,
the minimum of the priorities that occur infinitely often
in $\pi$,
is even.
A (memoryless) \emph{strategy} for player $a$ is a function $f_a \oftype V_a \to V$.
A play $\pi = v_0 v_1 \cdots$ is \emph{conform} to $f_a$ if for every $v_i \in \pi$, \;
$v_i \in V_a \impl v_{i+1} = f_a(v_i)$.
Player \Eloise is the \emph{winner} of the game if and only if there exists a winning strategy 
for player \Eloise, i.e., from the initial state every play conforming to the strategy will be
won by player \Eloise.

%

%% file: 03-lps2pbes.tex
\section{Translating modal \MUCALC and LPS to a modular PBES}\label{section:lps2pbes}

In this section we describe the adapted version of the translation from first order modal \MUCALC formulae and LPSs to PBESs.
The original translation for $\mu$CRL (of which \MCRLTWO is an extention) was published in \cite{groote2005:modelchecking}.
It translated the satisfaction of a formula by an LPS to a system of equations that consists
of one equation per propositional variable in the formula.
In order to be able to generate a parity game from the PBES efficiently, the equations
have to be partitioned in relatively independent parts.
In this section we present a translation that ensures that the structure of the equation system 
reflects the structure of the specification, for the modal operators in the formula.
This allows to partition the equations based on
the structure of the input specification.
To achieve that, we changed the original translation for the modal operators $\possibly{\alpha}$ and $\always{\alpha}$.
In our translation, for these operators new equations are introduced for every summand in the specification.

The translation is defined for a fixed process specification $P$, as defined in Def.~\ref{def:linear-process},
and for formulae of the form $\varphi_0 = \sigma \propvar{X}(x_f \oftype D_f \becomes d) \suchthat \varphi$, i.e., formulae with a fixpoint operator as outmost operator.

Satisfaction of the formula $\varphi_0$ by process $P$ is defined as $\E(\varphi_0)$, with the function $\E$ as defined below.

Let $\mathcal{D}$ be the set of data variables.
The function $\E \oftype \mathcal{M} \to \mathcal{E}^\ast$ 
generates a sequence of first order boolean equations for a formula $\varphi \in \mathcal{M}$.
The function $\E$ uses a function 
$\RHS \oftype \mathcal{M} \times \mathcal{D}^\ast \times \set{\nu, \mu} \to \PF \times \mathcal{E}^\ast$
(defined below) that produces the right hand side of the equation and a sequence of equations that are
introduced to be used in the right hand side. $\RHS(\varphi, \vec{v}, \varsigma)$ has a \MUCALC formula $\varphi$, a sequence of data variables $\vec{v}$ and a fixpoint operator $\varsigma$ as arguments. The latter two are needed for the newly introduced equations.

The function \E is specified as follows for the fixpoint operator:
\[
\E(\sigma \propvar{X}(x_f \oftype D_f \becomes d) \suchthat \varphi)
  \eqdef 
    \bigl( 
      \sigma \propvar{\fresh{X}}(\vec{v}) 
        = \psi
    \bigr)
    \concat Z
    \concat \E(\varphi) 
\]
with $\tuple{\psi, Z} = \RHS(\varphi, \vec{v}, \sigma)$
and $\vec{v} = [x_f \oftype D_f, x_p \oftype D_p]$. The operator $\concat$ denotes concatenation of sequences.
\\
A new equation is produced with $\psi$ as right hand side, which is the result of \RHS applied to the formula.
The resulting sequence of equations consists of the new equation together with $Z$, the equations generated by \RHS, and the result of \E applied
to the remainder of the formula, $\E(\varphi)$.
\\
For other operators, the function \E is applied to the subformulae recursively, where
the resulting sequences of equations are concatenated.

The function $\RHS$ generates the right hand sides of the equations as defined in Table~\ref{table:rhs}
\begin{table}[tp]
\caption{Definition of the \RHS function that generates a PBES from a \MUCALC formula, defined for a process $P$ as in Definition~\ref{def:linear-process}.}
\label{table:rhs}
\begin{tabular}{l@{\hspace{2pt}}l}
$\RHS(b, \vec{v}, \varsigma)$
  & $\eqdef \tuple{b, \emptyseq} $ \\

$\RHS(\neg b, \vec{v}, \varsigma)$
  & $\eqdef \tuple{\neg b, \emptyseq} $ \\
  
$\RHS(\varphi_1 \land \varphi_2, \vec{v}, \varsigma)$
  & $\eqdef \tuple{\psi_1 \land \psi_2, Z_1 \concat Z_2} $ 
 \quad with 
$\tuple{\psi_i, Z_i} = \RHS(\varphi_i, \vec{v}, \varsigma)\text{ for }i \in \set{1, 2}$.\\
  
$\RHS(\forall x \oftype D \suchthat \varphi, \vec{v}, \varsigma)$ 
  & $\eqdef \tuple{\forall x \oftype D \suchthat \psi, Z} $ 
\qquad with 
$\tuple{\psi, Z} = \RHS(\varphi, \vec{v}\concat[x], \varsigma)$ \\

$\RHS(\always{\alpha}\varphi, \vec{v}, \varsigma)$
  & $\eqdef \bigl\langle
  \propvar{\fresh{Y}}(\vec{v})
  ,$ \\
&\qquad
  $
  \langle\ \,
  \varsigma \propvar{\fresh{Y}}(\vec{v}) = 
    \Land_{i \in I} 
    \propvar{\fresh{X}_i}(\vec{v})
  ,$ \\
&\qquad\quad
  $  
  \varsigma \propvar{\fresh{X}_1}(\vec{v}) = 
    \ApplyGroup(1, \psi)
     $ \\
&\qquad\quad
  $\dotsb,$ \\
&\qquad\quad
  $
  \varsigma \propvar{\fresh{X}_m}(\vec{v}) = 
    \ApplyGroup(m, \psi)
     $ \\
&\qquad\quad
  $ Z   
  \; \rangle \; \bigr\rangle
  $ \\
& \quad with 
$\tuple{\psi, Z} = \RHS(\varphi, \vec{v}, \varsigma)$ \\
\end{tabular}

\begin{tabular}{l@{\hspace{2pt}}l}
$\RHS(\propvar{X}(d), \vec{v}, \varsigma)$ 
  & $\eqdef \langle \propvar{\fresh{X}}(d, x_p), \emptyseq \rangle $ \\
  
$\RHS(\sigma \propvar{X}(x_f \oftype D_f \becomes d) \suchthat \varphi, \vec{v}, \varsigma)$ 
  & $\eqdef \langle \propvar{\fresh{X}}(d, x_p), \emptyseq \rangle $
\end{tabular}
\end{table}
with the translation for the modal operator defined per summand $i$ as:
\[ \ApplyGroup(i, \psi) =
    \forall_{y \oftype E_i}
    (a_i(f_i(x_p,y)) \in \evaluation{\alpha}
     \land
     c_i(x_p,y))
   \impl \psi[x_p \becomes g_i(x_p,y)]
\]
The cases for boolean conditions $b$, negation and conjunction are straightforward.
For universal quantification the quantified variable is added to the list of parameters $\vec{v}$ for
equations generated by \RHS.
For propositional variables and fixpoint subformulae the corresponding variable is used,
where the notation $\propvar{\fresh{X}}$ is used to introduce a fresh variable that is guaranteed to be
unique in the equation system.\\
For the modal operator $\always{\alpha}\varphi$, the \RHS function
generates a new equation for each summand and results in an expression that is a 
conjunction of the propositional variables that are the left hand sides of these equations.
Every equation generated for a summand $i$ for a formula $\always{\alpha}\varphi$,
has a right hand side generated by $\ApplyGroup(i, \psi)$,
where $c_i$ is the guard of summand $i$, $g_i$ is the function that describes the transition
to the next state, $\psi$ is the result of applying \RHS to the formula $\varphi$, and
$a_i(f_i(x_p,y)) \in \evaluation{\alpha}$ means that the action of summand $i$ satisfies the 
action formula $\alpha$.\\
The cases for disjunction, existential quantification and possibility are similar to those for conjunction, 
universal quantification and necessity, respectively.

\begin{example}
For the two place buffer in Example~\ref{example:buffer2}, we want to verify that if a message $d$ is read through action `$\action{read}$', it will
 eventually be sent through action `$\action{send}$', which is expressed by the formula:
\[ \nu \propvar{Y} \suchthat
(\forall d \oftype D \suchthat (\always{\action{read}(d)}(\mu \propvar{X} \suchthat (\possibly{\ttrue}\ttrue \land \always{\neg\action{send}(d)}\propvar{X})))) 
\land \always{\ttrue}\propvar{Y} \]
Applying the function $\E$ results in the equation system in Figure~\ref{fig:example-pbes}.
\begin{figure}[tp]
\begin{tabular}{l@{}l@{\hspace{3pt}}l}
$\pbes$ & $\nu \propvar{Y}(q \oftype \container{List}(D)) $ 
        & $= (\forall d \oftype D \suchthat (\#q < 2) \impl \propvar{X}(q \append d, d)) $
          $\land \propvar{Y_1}(q);$ \\
        & $\nu \propvar{Y_1}(q \oftype \container{List}(D)) $ 
        & $= \propvar{Y_{11}}(q) \land \propvar{Y_{12}}(q);$ \\
        & $\nu \propvar{Y_{11}}(q \oftype \container{List}(D)) $ 
        & $= (\forall d' \oftype D \suchthat (\#q < 2) \impl \propvar{Y}(q \append d')) $\\
        & $\nu \propvar{Y_{12}}(q \oftype \container{List}(D)) $ 
        & $= ((q \neq \emptyseq) \impl \propvar{Y}(\tail(q)));$ \\
        & $\mu \propvar{X}(q \oftype \container{List}(D), d\oftype D) $ 
        & $= \propvar{X_1}(q) \land \propvar{X_2}(q, d);$ \\
        & $\mu \propvar{X_1}(q \oftype \container{List}(D)) $ 
        & $= \propvar{X_{11}}(q) \lor \propvar{X_{12}}(q);$ \\
        & $\mu \propvar{X_{11}}(q \oftype \container{List}(D)) $ 
        & $= (\#q < 2)$ \\
        & $\mu \propvar{X_{12}}(q \oftype \container{List}(D)) $ 
        & $= (q \neq \emptyseq)$ \\
        & $\mu \propvar{X_2}(q \oftype \container{List}(D), d\oftype D) $ 
        & $= \propvar{X_{21}}(q, d) \land \propvar{X_{22}}(q, d);$ \\
        & $\mu \propvar{X_{21}}(q \oftype \container{List}(D), d\oftype D) $ 
        & $= (\forall d' \oftype D \suchthat (\#q < 2) \impl \propvar{X}(q \append d', d)) $\\
        & $\mu \propvar{X_{22}}(q \oftype \container{List}(D), d\oftype D) $ 
        & $= (\head(q) = d \lor q = \emptyseq \lor \propvar{X}(\tail(q), d));$ \\
$\init$ & $\propvar{Y}(\emptyseq);$ 
\end{tabular}
\caption{Example PBES.}
\label{fig:example-pbes}
\end{figure}
The equation $\propvar{Y}$ represents the $\nu \propvar{Y}$ part of the formula (which is the whole formula), 
$\propvar{X}$ represents the $\mu \propvar{X}$ part.
The system is initialised to the toplevel variable ($\propvar{Y}$) with the LPS parameters set to
their initial value in the specification ($q = \emptyseq$).
\\
The equation $\propvar{Y_1}$ represents the ``$\always{\ttrue}\propvar{Y}$'' part (at the end) of the formula, a conjunction of $\propvar{Y_{11}}$ and $\propvar{Y_{12}}$, each representing
a set of transitions from one of the summands of the LPS, followed by $\propvar{Y}$ with
the parameters updated to reflect the new state after the transitions.
The equation $\propvar{X_1}$ represents ``$\possibly{\ttrue}\ttrue$'' (`some action is enabled'),
 a disjunction of $\propvar{X_{11}}$ and $\propvar{X_{12}}$,
representing that of one of the two summands is enabled.
The equation $\propvar{X_2}$ represents ``$\always{\neg\action{send}(d)}\propvar{X}$'' (`\propvar{X} should hold after every action other than $\action{send}(d)$'),
 a conjunction of $\propvar{X_{21}}$ and $\propvar{X_{22}}$,
representing the transitions that match the action formula $\neg\action{send}(d)$.
\end{example}

Having separate equations representing the different summands now allows
the tools to make a partition based on the summands, thus reflecting the structure of the LPS.

%% file: 04-spg-generation.tex
\section{Generating Symbolic Parity Games}\label{section:spg-generation}

Instantiation to a parity game is similar to generating the reachable state space from a specification; both involve generating a large graph out of an abstract description.
That is the reason we implemented instantiation of PBESs to parity games as an extension
of 
the high performance verification toolset \LTSMIN\cite{blom2010:ltsmin}. \LTSMIN is modular in the sense that the core algorithms are separated from the input languages
by using a generic interface in between.
We extended \LTSMIN
with a PBES language module for generating symbolic parity games.
We discuss instantiation in this section briefly. 
Details on the language module, the dependency matrix for PBESs and instantiation to parity games can be found in \cite{kant2012:efficient}.

The result of instantiation is a symbolic parity game, 
in which MDDs are used to represent the set of reachable states of the game, for each player and
for each priority the set of states owned by that player respectively with that priority, and 
the relations encoding the moves, partitioned in transition groups.

\subsection{Parameterised Parity Games}

For the instantiation of PBESs to Parity Games, we assume PBESs to be in a specific form: the \emph{Parameterised Parity Game} (PPG), an equation system where every equation is either conjunctive or disjunctive.
Not every PBES generated from an LPS and a formula (by the translation in Section~\ref{section:lps2pbes}) is a PPG, 
but any PBES can be transformed to an equivalent PPG by a transformation described in \cite{kant2012:efficient}.
During instantiation every variable associated with a conjunctive expression will
belong to player \Abelard, variables with disjunctive expressions will be owned by
player \Eloise.

\subsection{State Vectors and the Partitioned Transition Relation}\label{section:partitioning}

Instantiated variables (predicate variables with concrete parameters) are the \emph{states} in the generated parity game.
A variable and its data parameters are stored in what in \LTSMIN is called a \emph{state vector}. 
State vectors are used to encode states in LTSs and parity games, and are
vectors of integers 
$\tuple{x_0, x_1, \cdots, x_K}$ (other value types are stored in a database).

Logical dependencies as expressed by the equations are encoded as \emph{transitions} in 
the generated parity game and are
computed by a successor function \nextst.

During generation, \LTSMIN builds a symbolic transition relation $E$ from the 
transitions that are computed by the language module. 
In \LTSMIN, $E$ is a partitioned transition relation $E = E_1 \cup \ldots \cup E_M$, 
consisting of parts $E_g$ which are called \emph{transition groups}.
The parts $E_g$ are stored as MDDs; the composite relation $E$ is not stored. 
For every newly encountered state its successors are computed for every transition group.

Applying the 
partitioned transition relation can be much more efficient than
applying a monolithic transition relation if the partition is chosen well, as
is well known in the literature
(see, e.g., \cite{burch1994:symbolic}, \cite{ranjan1995:efficient}).

Computing the successors of a set of states $V$ is then defined as 
$\nextst(V) = \bigcup_{1 \leq g \leq M} \nextst_g(V)$,
i.e., iterating over the transition groups, where $\nextst_g(V)$ is the result
of applying the relation $E_g$ to $V$ (and renaming the variables).

It is known that the order in which the different parts of the transition relation
are applied often does matter for performance. 
For instance, saturation \cite{ciardo2001:saturation} is
a technique for optimising the order of application to minimise 
the size of the intermediate decision diagrams.
In \LTSMIN several of such techniques are available.

There are several ways of choosing transition groups for PPGs; we distinguish two.
First, choosing entire equations to form a transition group (which we call \emph{simple}).
This approach is best when the equation system is generated by the translation in Section~\ref{section:lps2pbes}:
if the summands of the original LPS are relatively independent, then also the equations in the PBES will be relatively independent.
Second, splitting conjunctive equations in conjuncts and disjunctive equations in disjuncts
(which we call \emph{splitting}) is an option when such independence is not present in the equation system (e.g., when using the previous \texttt{lps2pbes} translation). 
However, as will be demonstrated in the experiments, splitting conjuncts and disjuncts does not per se result in a good partition.

\subsection{Dependency matrix}

On top of the partitioning of the transition relation into \emph{transition groups}, we use a 
\emph{dependency matrix} to store information about the \emph{dependence} of transition groups on the \emph{parts} $x_i$ of the state vector. 
A group $g$ is \emph{dependent} on part $i$ if the variable that is stored in slot $i$
is read or changed by the expression of group $g$.
Independence is also referred to as \emph{locality}: when the dependency matrix is sparse, the effect of transition groups is relatively local.
A detailed description is in \cite{kant2012:efficient}, we will explain it here using an example.

\begin{example}
Suppose we have an LPS that models the Tic Tac Toe game as a process with parameters
$b_{1}$ \ldots $b_{9}$ to encode the board configuration and parameter $p$ to 
encode whose turn it is. 
Suppose that the LPS has separate summands for every position on the board, i.e.,
a summand for placing a piece on position $b_{1}$, one for $b_{2}$, etc.
We want to check the property that player $X$ has a winning strategy: 
$\;
\mu \propvar{Z} \suchthat \always{\action{wins}(O)}\tfalse \land \possibly{\action{move}(X)}(\possibly{\action{wins}(X)}\ttrue \lor \always{\action{move}(O)}\propvar{Z})
$,\;
which for clarity of the example we present as a modal equation system:
\begin{align*}
\mu \propvar{Z} & = \propvar{A} \land \propvar{B} &
\mu \propvar{B} &= \possibly{\action{move}(X)} \propvar{C} \\
\mu \propvar{A} &= \always{\action{wins}(O)}\tfalse &
\mu \propvar{C} &= \possibly{\action{wins}(X)}\ttrue \lor \always{\action{move}(O)}\propvar{Z} \enspace .
\end{align*}
The resulting PBES for this property is:
\begin{quote}
\begin{tabular}{l@{}l@{\hspace{3pt}}l}
$\pbes$ & $\mu \propvar{Z}(b_{1}, \dotsc, b_{9}, p)$ 
        & = $\propvar{A}(b_{1}, \dotsc, b_{9}, p) \land \propvar{B}(b_{1}, \dotsc, b_{9}, p); $ \\
 & $\mu \propvar{A}(b_{1}, \dotsc, b_{9}, p)$ 
        & $ = \propvar{A_{1}}(b_{1}, \dotsc, b_{9}, p) \land \propvar{A_{2}}(b_{1}, \dotsc, b_{9}, p) \land \dotsb $ \\
        & \ldots \\
 & $\mu \propvar{B}(b_{1}, \dotsc, b_{9}, p)$ 
        & $ = \propvar{B_{1}}(b_{1}, \dotsc, b_{9}, p) \lor \propvar{B_{2}}(b_{1}, \dotsc, b_{9}, p) \lor \dotsb $ \\
 & $\mu \propvar{B_{1}}(b_{1}, \dotsc, b_{9}, p)$ 
        & $ = (b_{1}=-) \land \propvar{C}(p, b_{2}, \dotsc, b_{9}, \mathit{Opponent}(p))$ \\
 & \ldots \\ 
 & $\mu \propvar{B_{9}}(b_{1}, \dotsc, b_{9}, p)$ 
        & $ = (b_{9}=-) \land \propvar{C}(b_{1}, \dotsc, b_{8}, p, \mathit{Opponent}(p))$ \\                
& $\mu \propvar{C}(b_{1}, \dotsc, b_{9}, p)$ 
        & $ = $ \ldots \\
        & \ldots \\

$\init$ & \multicolumn{2}{l}{$\propvar{Z}(-, -, -, -, -, -, -, -, -, \text{X});$} \\
\end{tabular}
\end{quote}
\end{example}
\noindent Equation $\propvar{A}$ is a conjunction of the equations $\propvar{A_i}$ for $1 \leq i \leq 9$, 
where each equation $\propvar{A_i}$ means that $\always{\action{wins}(O)}\tfalse$ holds for summand $i$ in the LPS; 
in other words, that there is no action $\action{wins}(O)$ enabled in that summand.
Equation $\propvar{B}$ is a disjunction of the equations $\propvar{B_i}$ for $1 \leq i \leq 9$, 
where each equation $\propvar{B_i}$ means that $\possibly{\action{move}(X)}\propvar{C}$ holds for summand $i$; 
in other words, that there is an action $\action{move}(X)$ enabled, which is true if $b_i = -$, 
and that afterwards $\propvar{C}$ holds for the state resulting from the action.

Each of $\propvar{B_{1}}$ to $\propvar{B_{9}}$ represents a single move on the board,
and is computed by a single transition group touching only a small
number of parameters.

\begin{wrapfigure}{r}{0.38\textwidth}
\vspace{-20pt}
\begin{center}
\scalebox{.9}{
\begin{tabular}{l|ccccccc}
$g$ & $\propvar{Var}$ & $b_{1}$ & $b_{2}$ & $b_{3}$ & \ldots & $b_{9}$ & $p$\\
\hline
\propvar{Z}&$+$&$-$&$-$&$-$& &$-$ &$-$\\
\propvar{A}&$+$&$-$&$-$&$-$& &$-$ &$-$\\
\ldots & & & & & & & \\
\propvar{B}&$+$&$-$&$-$&$-$& &$-$&$-$\\
\propvar{B_{1}}&$+$&$+$&$-$&$-$& &$-$&$+$\\
\propvar{B_{2}}&$+$&$-$&$+$&$-$& &$-$&$+$\\
\ldots & & & & & & & \\
\propvar{B_{9}}&$+$&$-$&$-$&$-$& & $+$&$+$\\
\propvar{C}&$+$&$-$&$-$&$-$& &$-$&$-$\\
\ldots & & & & & & & \\
\end{tabular}
}
\end{center}
\vspace{-10pt}
\caption{Dependency matrix for the Tic Tac Toe game.}
\label{fig:dm}
\vspace{-20pt}
\end{wrapfigure}

The resulting dependency matrix is in Figure~\ref{fig:dm}.
For the transition groups $\propvar{B_{1}}$ to $\propvar{B_{9}}$,
repectively the board parameters $b_{1}$ to $b_{9}$ are marked as dependent ($+$).
The group of equation $\propvar{Z}$ only changes the predicate variable $\propvar{Var}$
and none of the parameters.
This way the matrix is very sparse and transitions can be encoded efficiently.

If a transition group $g$ is only dependent on, for instance, parameters 1 and 3 (as for $\propvar{B_{2}}$
in the example), then a transition
$\;
\tuple{\mathbf{0}, 0, \mathbf{1}, 1, 1, \dotsc} \to_g \tuple{\mathbf{2}, 0, \mathbf{2}, 1, 1, \dotsc}
\;$
is simply stored as a vector of tuples of old and new values for dependent parameters:
$\tuple{\tuple{0,2},\tuple{1,2}}$
Also, once a state has been visited with values $\tuple{0,2}$ for parameters 1 and 3, for future
fresh states with the same values for these parameters, transitions for group $g$ do not have
to be computed for that state again.
The MDDs used to store the partitioned transition relation only use these shorter
vectors of integers, allowing for a compact representation of the transition relation.

%% file: 05-spg-solver.tex
\section{Symbolic Parity Game Solver}\label{section:spg-solver}

We implemented the recursive algorithm for solving parity games by Zielonka \cite{zielonka1998:infinite}
for symbolic parity games.
The recursive algorithm is widely used in practice and easy to implement symbolically.
Although its worst-case complexity is worse than some other algorithms, its performance is very good in practice \cite{friedmann2009:solving} (at least for explicit representations of the game).
Our solver is similar to the symbolic parity game solver by Bakera et al. \cite{bakera2009:solving}.
Our own implementation allows us to reuse the partitioned transition relation from the instantiation tool directly.

The algorithm returns the set of winning states for player \Abelard and the set of winning states for player \Eloise
as MDDs.
%
%
The algorithm makes heavy use of the successor function \nextst\xspace and predecessor function \prevst, 
which use the partitioned transition relation in the tool.
However, in the solver currently available in \texttt{pbes2lts-sym} no saturation or similar techniques are used;
the relations of transition groups are always applied in the order $1 .. M$.


%% file: 06-experiments.tex
\section{Experiments}\label{section:experiments}

We performed experiments to compare our new tool to existing methods for solving
PBESs (available in the \MCRLTWO toolset), previous versions of our tool and \NuSMV 2 \cite{cimatti2002:nusmv2}.

As benchmarks we verified properties for two models. 
First, we created a model of the well known \emph{Connect Four} game (\texttt{four}) in both \MCRLTWO and \NuSMV
with different board sizes (the original is 7$\times$6). 
We verified whether player Yellow has a winning strategy:
$ \mu \propvar{X} \suchthat \always{\action{wins}(\mathit{Red})}\tfalse 
  \land \possibly{\action{move}}(\possibly{\action{wins}(\mathit{Yellow})}\ttrue \lor \always{\action{move}} \propvar{X})
$.
For \NuSMV, we used an SMV model and a CTL property equivalent to the \MUCALC property:
$ 
\mathbf{EX} (\action{yellowwins} \lor \mathbf{AX} (\neg\action{redwins} \land 
\mathbf{EX} (\action{yellowwins} \lor \mathbf{AX} ( \ldots ) ) ) ) 
$.

Second, we verified several properties for \emph{Finite State Machines} (FSMs) that are used in the 
Compact Muon Solenoid (CMS) detector, part of the Large Hadron Collider (LHC) at CERN. 
These state machines are used to control all the components of the detector, which are organised in a hierarchical manner. Components send commands to their children, which send status updates to their parent, asynchronously.
In the experiments, we used the \texttt{wheel} subsystem, consisting of 8 FSMs, which we checked for four properties:
absence of deadlocks (\texttt{nodeadlock}),
absence of intermediate states in the \emph{when} phase (\texttt{absence}),
\texttt{progress} and \texttt{responsiveness}.
The FSMs, the translation from FSMs to \MCRLTWO, and the properties have been reported in \cite{hwong2013:formalising}.
We did not compare to \NuSMV for this model.

The \MCRLTWO models are translated to PBESs using the function $\E$, 
described in Section~\ref{section:lps2pbes}, which preserves the structure of the LPS (\texttt{structured}),
and using the unstructured earlier version of the translation (\texttt{unstructured}).
Both are available in the \MCRLTWO toolset.
The structured version can be used by passing the \texttt{-s} option to the \texttt{lps2pbes} tool.
We compared three tool combinations\footnote{Details on versions of the tools and options passed to the tools can be found on \url{http://www.cs.utwente.nl/\~kant/graphite2014/}.}: 
\begin{itemize}
\item \texttt{pbes2bool}, one of the explicit state PBES solvers in the \MCRLTWO toolset, which instantiates to a Boolean Equation System (BES) and solves the BES using approximation. 
We used the unstructured \texttt{lps2pbes} as the tool performed better with that translation.
\item The \LTSMIN toolset -- in particular the tool \texttt{pbes2lts-sym} in combination with our new symbolic parity game solver \texttt{spgsolver}.
We compared three combinations of \texttt{lps2pbes} translations and transition partitioning (see Section~\ref{section:partitioning}):
\texttt{simple}: unstructured \texttt{lps2pbes} with one group per equation;
\texttt{split}: unstructured \texttt{lps2pbes} with the equations split into conjuncts/disjuncts;
and \texttt{structured}: structured \texttt{lps2pbes} with one group per equation.
\item \NuSMV 2.5.4 with the \texttt{-dynamic} option to enable dynamic reordering of variables,
which appeared to give better performance.
\end{itemize}

The experiments were performed on a machine with two quad-core Intel Xeon E5520 CPUs @ 2.27 GHz and 24GB
memory. Every tool was given a 20 GB memory limit and a 24 h time limit. 
We report the number
of states of the generated symbolic parity game, the number of MDD nodes used to store the set of states and the number of MDD nodes used for storing the relations.

\begin{table}
\caption{Experimental results for ConnectFour and the CERN case study.
Time is measured in seconds, memory usage in multiples of 1,000 KiB.
`gen' indicates time and memory used for generating a parity game,
`solve' solving and `total' and `max' indicate the total time and 
maximum memory used in all steps combined.}
\label{table:results}
\begin{center}
\scalebox{0.76}{
\begin{tabular}{ll@{\hspace{5pt}}r@{\hspace{5pt}}r@{\hspace{5pt}}r@{\hspace{5pt}}rrr@{\hspace{5pt}}rrr}
 & 
 &
\multicolumn{1}{c|}{} &
\multicolumn{1}{c|}{MDD} &  
\multicolumn{1}{c|}{Trans.} & 
\multicolumn{3}{c|}{Time (s)} & 
\multicolumn{3}{c}{Memory ($\times$1,000 KiB)} \\
System & 
Tool &
\multicolumn{1}{c|}{\#States} &
\multicolumn{1}{c|}{nodes} &
\multicolumn{1}{c|}{nodes} &
gen & solve & \multicolumn{1}{c|}{total} &
gen & solve & max \\
\toprule
\texttt{four.5x4}
 & \NuSMV
 &  &  &
 &  & & 74 
 & & & 67     \\
 & \texttt{pbes2bool}
 & $1.8 \cdot 10^6$ &        &
 & & & 118
 & & & 1,678 \\
 & \LTSMIN \texttt{simple} 
 & $1.3 \cdot 10^7$ & $5.4 \cdot 10^5$       & $2.0 \cdot 10^6$
 &  1,136 & 171 & 1,307
 & 215 & 298 & 298 \\
 & \LTSMIN \texttt{split}
 & $2.1 \cdot 10^7$ & $8.5 \cdot 10^4$       & $3.1 \cdot 10^3$
 &   19 & 84 & 103 
 & 52 & 118 & 118   \\
 & \LTSMIN \texttt{struct}
 & $1.3 \cdot 10^7$ & $4.5 \cdot 10^4$       & $6.6 \cdot 10^2$
 &      7.3 & 37 & 44 
 & 49  & 76 & 76    \\
\midrule
\texttt{four.6x4}
 & \NuSMV
 &  &   &
 &  & & 2,657 
 & & & 432    \\
 & \texttt{pbes2bool}
 &  &   &
 &  & & -- & &  & $>$20,000     \\
 & \LTSMIN \texttt{split}
 & $ 7.1 \cdot 10^{8}$ & $ 2.8 \cdot 10^5$       & $4.3 \cdot 10^3$
 & 94   & 1,435 & 1,529  
 & 65  &  1,242 &  1,242  \\
 & \LTSMIN \texttt{struct}
 & $ 3.8 \cdot 10^{8}$ & $ 1.5 \cdot 10^5$       & $8.9 \cdot 10^2$
&  13   & 565 & 586   
&  50  & 770 &  770     \\
\midrule
\texttt{four.6x5}
 & \NuSMV
 &  &   &
 &  & & $>$86,400 & &  & --    \\
 & \LTSMIN \texttt{split}
 & $4.6 \cdot 10^{10}$ & $1.8 \cdot 10^6$       & $6.3 \cdot 10^3$
 & 949   &  37,697 & 38,646
 & 328    & 13,706    & 13,706   \\
 & \LTSMIN \texttt{struct}
 & $2.6 \cdot 10^{10}$ & $1.6 \cdot 10^6$       & $1.3 \cdot 10^3$
 &      574 & 17,139  & 17,713
 & 324  &  13,706 & 13,706    \\
\midrule
\texttt{four.7x5}
 & \NuSMV
 &  &  &
 &  & & $>$86,400 & &  & --    \\
 & \LTSMIN \texttt{split}
 & $3.1 \cdot 10^{12}$ & $5.1 \cdot 10^6$ & $8.4 \cdot 10^3$
 & 6,988 & -- & --  
 & 1,300 & $>$20,000 & $>$20,000   \\
 & \LTSMIN \texttt{struct}
 & $1.6 \cdot 10^{12}$ & $3.3 \cdot 10^{6}$ & $1.7 \cdot 10^3$
 & 1,122 & -- & --
 & 520 &  $>$20,000 & $>$20,000   \\
\midrule
\texttt{four.6x6}
 & \LTSMIN \texttt{struct}
 & $1.6 \cdot 10^{12}$ & $2.2 \cdot 10^{7}$ & $1.8 \cdot 10^3$
 & 10,750 & -- & --     
 & 5,439 & $>$20,000 &$>$20,000   \\
\midrule
\texttt{four.7x6}
 & \LTSMIN \texttt{struct}
 & $2.0 \cdot 10^{14}$    &  $7.2 \cdot 10^{7}$ & $2.2 \cdot 10^3$
 & 81,116     & $>$86,400  & $>$86,400    
 & 14,340   & $>$20,000 & $>$20,000 \\
\midrule
\texttt{wheel}
 & \texttt{pbes2bool}
 & $4.6 \cdot 10^6$  &   &
 &   &   & 2,190   
 &   &   & 9,894 \\
\texttt{nodeadlock}
 & \LTSMIN \texttt{split}
 & $4.6 \cdot 10^6$  & $2.7 \cdot 10^5$ & $2.6 \cdot 10^6$
 & 16,856  & 247  &  17,103 
 & 340  & 187  & 340 \\
 & \LTSMIN \texttt{struct}
 & $1.4 \cdot 10^7$  & $3.1 \cdot 10^5$ & $1.2 \cdot 10^4$
 & 190  & 47  & 237 
 & 98  & 118   & 118 \\
\midrule
\texttt{wheel}
 & \texttt{pbes2bool}
 & $5.9 \cdot 10^6$ &   &
 &   &   & 4,420  
 &   &  & 14,779 \\
\texttt{absence}
 & \LTSMIN \texttt{split}
 & $2.0 \cdot 10^7$  & $1.2 \cdot 10^6$ & $1.1 \cdot 10^7$
 & 28,769  & 477  & 29,246 
 & 889  &  770   & 889 \\
 & \LTSMIN \texttt{struct}
 & $2.4 \cdot 10^7$  & $8.5 \cdot 10^5$  & $1.3 \cdot 10^4$
 & 1,142  &  172 & 1,314 
 & 215  & 118    & 215 \\
\midrule
\texttt{wheel}
 & \texttt{pbes2bool}
 &  &   &
 &   &   & --   
 &   &  & $>$20,000  \\
\texttt{progress}
 & \LTSMIN \texttt{split}
 & $3.2 \cdot 10^7$  & $1.4 \cdot 10^7$ & $1.3 \cdot 10^8$
 & 61,156  &  9,078  & 70,234  
 & 9,050  & 8,474  & 9,050 \\
 & \LTSMIN \texttt{struct}
 & $6.5 \cdot 10^7$  & $3.1 \cdot 10^5$  & $1.2 \cdot 10^4$
 & 1,266  &  2,471  & 3,737
 & 142  & 299  & 299 \\
\bottomrule
\end{tabular}
}
\end{center}
\end{table}

\subsection{Results}
The results are in Table~\ref{table:results}.
The models are ordered by their number of states. When a tool was not able to complete within the constraints,
the larger model was skipped for that tool.
For the \texttt{responsiveness} property the \texttt{pbes2bool}
did not complete within the 20 GB memory bound and \texttt{pbes2lts-sym} could
not generate the parity game within 24 h. Also, \texttt{four.7x6} could not be solved by any of the tools.

We can make the following observations.
From the results we see that of the different options for \texttt{pbes2lts-sym},
our new approach performs always best, both in time and memory, compared to the splitting
and simple approach.
For both Connect Four and the Wheel model, \texttt{pbes2bool} is up to 9 times slower and uses up to more than 80 times more memory than the structured approach with \texttt{pbes2lts-sym}, for cases
where both tools finished within the constraints.
Comparison with \NuSMV shows a mixed picture. Structured \texttt{pbes2lts-sym} is up to 4.5 times faster, but \NuSMV uses up to 1.9 times less memory (considering the structured approach).
Also, \NuSMV exceeds the time limit for \texttt{four.6x5} and larger and \texttt{pbes2lts-sym}
exceeds the memory limit for \texttt{four.7x5} and larger. 
That \NuSMV is slower can be explained by the the conjunctive partitioning that is used in the tool,
while the non-determinism in the Connect Four gives rise to a disjunctive specification, which
is exploited by the disjunctive partitioning in \texttt{pbes2lts-sym}.
The better memory performance of \NuSMV is due to the encoding of the board using two bits per field,
where \texttt{pbes2lts-sym} uses a 32-bit integer for every field.

There is a large difference between the different models in that for the Connect Four game most time and memory is spent on solving the generated game, not on generating it; for the \texttt{wheel} FSMs it is the other way around.

%% file: 07-conclusions.tex
\section{Conclusions}

We have presented an improved method for verifying modal \MUCALC for process algebraic
specification,
consisting of an improved translation of the verification problem to a PBES (\texttt{lps2pbes}),
an efficient tool for symbolic instantiation of the PBES to a symbolic parity game
based on \LTSMIN (\texttt{pbes2lts-sym}),
and a new symbolic parity game solver tool (\texttt{spgsolver}).
The combination of these tools allows for high performance model checking of large systems,
using MDDs as data structures. The structure of the specification is used for choosing
a good partition of the transition relation, allowing for efficient application of operations 
on these data structures.

We compared the performance of the new solution to the existing tool \texttt{pbes2bool}
and the symbolic model checker \NuSMV. 
The new \LTSMIN based tools perform much better than the previous version and than
\texttt{pbes2bool} both in execution time and memory usage.
\NuSMV is more memory efficient, but slower, in comparison to our approach. 

We intend to experiment with combinations of disjunctive and conjunctive
partitioning in the \texttt{pbes2lts-sym} tool and the symbolic parity game solver.
We want to extend \LTSMIN to allow for translation to a parity game for any supported input language.
Furthermore, we want to apply optimisations like saturation and the parallel application of MDD operations in the parity game solver.